\documentclass{PoS}
\title{Asymmetric CFTs and GSUGRA}

\ShortTitle{Asymmetric CFTs and GSUGRA}

\author{\speaker{Ralph Blumenhagen}, \speaker{Michael Fuchs}\\
        Max-Planck-Institut f\"ur Physik (Werner-Heisenberg-Institut), \\ 
   F\"ohringer Ring 6,  80805 M\"unchen, Germany \\
        E-mail: \email{blumenha,mfuchs@mpp.mpg.de}}

\author{Erik Plauschinn\\
        Arnold-Sommerfeld-Center f\"ur Theoretische Physik \\
        Ludwig-Maximilians-Universit\"at M\"unchen \\ 
        Theresienstra\ss e 37, 80333 M\"unchen, Germany\\
        E-mail: \email{erik.plauschinn@lmu.de}}

\abstract{Recent progress on the relation between asymmetric conformal field theories
and vacua of gauged supergravities is reviewed. This includes a classification of
asymmetric Gepner models in 8D, 6D and 4D with at least eight supercharges, and how
they can be categorized by just a few mechanisms  including the super Higgs-effect.  
The latter is a prerequisite for the identification with non-geometric flux compactifications.
We also highlight  our findings for the  identification of  4D asymmetric Gepner models
with ${\cal N}=1$ supersymmetry with Minkowski vacua  of ${\cal N}=2$ GSUGRA.
}

\FullConference{Corfu Summer Institute 2016 ``School and Workshops on Elementary Particle Physics and Gravity''\\
31 August - 23 September, 2016\\
Corfu, Greece}

\usepackage{amsmath}
\usepackage{amsfonts}
\usepackage{amssymb}
\usepackage{graphicx}

\newcommand{\eq}[1]{\begin{equation}
                     \begin{split} #1 \end{split}
                     \end{equation}}

\newcommand{\ov}{\overline}

\allowdisplaybreaks[2]
\numberwithin{equation}{section}

\begin{document}


\section{Introduction}

Two approaches are frequently used to study string theory:
The first utilizes a world-sheet description of the string, given by a $2D$ conformal field theory (CFT), with  the advantage that CFTs are exact (classical) solutions to the string equations of motion including $\alpha'$-corrections. Disadvantages are that loop corrections and D-brane instantons have to be added by hand, and that a description of non-trivial Ramond-Ramond (R-R) backgrounds is not known in general. Furthermore, given a particular CFT it is usually hard to identify the corresponding target-space geometry, and many CFTs are believed to even lack a smooth target space, especially when (asymmetric) orbifold techniques \cite{Dixon:1985jw,Dixon:1986jc}  are used to stabilize moduli.

The second approach to studying string theory uses effective field theory, i.e. a supergravity, and therefore takes a target space viewpoint. It captures only the leading effects in $\alpha '$ such that it is not guaranteed that solutions lift to full string solutions. Having a solution of the supergravity, usually many moduli stay unstabilized. To stabilize them one usually perturbs the solution with additional fluxes in the internal geometry. Clearly, the fluxes have to be small and their backreaction onto the geometry must be negligible.

From the supergravity viewpoint the  fluxes can induce the gauging of certain isometries of the scalar manifold, resulting in a gauged supergravity (GSUGRA). But from several angles like T-duality \cite{Shelton:2005cf}, 
$SU(3) \times SU(3)$-structure compactifications \cite{Grana:2005ny,Grana:2006hr}, 
double field theory \cite{Hull:2009mi,Aldazabal:2013sca} or the embedding tensor formalism \cite{Samtleben:2008pe,Trigiante:2016mnt} it is clear that there are more gaugings than fluxes from the internal geometry. Since they cannot come from the internal geometry they are in turn called non-geometric gaugings/fluxes. For instance, on a $T^3$ they are connected to the geometric ones by a T-duality in two or  three directions. Therefore the non-geometric fluxes correspond to the winding duals of the geometric fluxes and transcend the usual supergravity description.
In fact their natural home is double field theory, which is a manifest $O(D,D)$ 
covariant field theory, where the usual coordinates are extended by  so-called winding coordinates.
There are two  concerns about such non-geometric fluxes. 
As yet it is not clear, which (non-geometric) fluxes can be turned on simultaneously. The strong constraint
in DFT seems to be stronger than the quadratic constraints of GSUGRA.  Second,  it was shown in \cite{Blumenhagen:2015kja} that the backreaction of non-geometric fluxes is in general of order one and therefore at the boundary  of  control. 

The question is therefore, whether the non-geometric gaugings of supergravity are  part of the string landscape or rather of the swampland. An affirmative answer to this question could be given if one could construct the corresponding conformal field theory, thus the full string solution, to such a background. 

A hint how this CFT could look like can be inferred by noting that the non-geometric fluxes are the T-duals of the geometric ones. Since T-duality corresponds to a reflection of the right-moving coordinates on the worldsheet, it is an intrinsically asymmetric operation on the worldsheet. As such it is natural to look for asymmetric CFTs. Beginning with asymmetric orbifolds \cite{Narain:1986qm}, many asymmetric CFTs were constructed \cite{Antoniadis:1985az,Hellerman:2002ax,Flournoy:2004vn,Flournoy:2005xe,Hellerman:2006tx}. In the context of asymmetric orbifolds of torus compactifications also a link to the flux algebra was established \cite{Condeescu:2012sp,Condeescu:2013yma}. We asked ourselves how far this link might go: Is there a  connection between asymmetric conformal field theories and gaugings in supergravity? Could it be, that these two seemingly different approaches to moduli stabilization are in fact the target space and worldsheet perspectives of the same set of vacua?

In the first  paper \cite{Blumenhagen:2016axv}, we analyzed this question in the context of Calabi-Yau compactifications and in a second paper \cite{Blumenhagen:2016rof}, we confirmed the correspondence in more controlled setups with  more supersymmetry. The  results are encouraging and suggest that there is indeed a deep correspondence between asymmetric CFTs and gauged supergravities. 
Recalling the problem with the large backreaction in presence of non-geometric gaugings, the precise statement we make is therefore: We conjecture a correspondence between certain asymmetric CFTs and fully backreacted solutions of gauged supergravity. 
If correct, this has interesting consequences: 
\begin{itemize}
\item Some Minkowski vacua  of gauged supergravity do lift to full string vacua. 
\item Since some breakings in supergravity are only allowed when geometric and non-geometric gaugings are turned on simultaneously, non-geometric gaugings are indeed part of the string landscape. 
\item Partial supersymmetry breaking is possible at all  orders in $\alpha'$.
\end{itemize}

This proceedings article is organized as follows: In  section 2 we review the essential ingredients used in our construction. We start with simple currents, continue with Gepner models, the construction of Calabi-Yau manifolds and close with a review of gauged supergravities and especially their supersymmetry breaking mechanisms. In section~\ref{sec_corr} we review \cite{Blumenhagen:2016axv} and \cite{Blumenhagen:2016rof}. We start with the case of $4D$ ${\cal N}=2$ to ${\cal N}=1$ supersymmetry breaking which we analyzed in \cite{Blumenhagen:2016axv}. Then we present parts of our full classification of asymmetric Gepner models in any dimension with extended supersymmetry to give more evidence for our conjecture as done in \cite{Blumenhagen:2016rof}.

We also would like to highlight the papers \cite{Israel:2013wwa,Israel:2015efa}, in which asymmetric 
Gepner models and their relation to Calabi-Yau manifolds were  studied. Here we go beyond this analysis and provide a connection between asymmetric Gepner models and gauged supergravity theories.

\section{Preliminaries}
In this section we  review the main ingredients of our construction: Simple currents, Gepner models and gauged supergravity theories. We  refrain from being too technical and focus on a heuristic understanding of the mechanisms we employ. 

\subsection{Simple Currents}
Let us first recall the basic facts about partition functions in conformal field theories (CFT) (for an introduction see e.g. \cite{Blumenhagen:2009zz} or \cite{Blumenhagen:2013fgp}). In string theory, the one-loop diagram is a worldsheet with the topology of a torus. As usual all particles can run in the loop but their contribution depends on their mass. Evaluating the one-loop vacuum amplitude gives rise to the partition function $Z$ which counts the number of states at each energy level and collects them in a power series. In closed string theory, every target space field is a combination of a left- and a right-moving worldsheet piece. Having this in mind it is clear that the partition function $Z$ can be written as
\eq{
Z(\tau, \ov \tau) = \sum_{i,j} \chi_i(\tau) \, M_{ij} \, \chi_j(\ov \tau) \,,
}
where $M_{ij}$ is a matrix which dictates, in which way the left- and right-moving modes (thus holomorphic and antiholomorphic) pieces may be combined to give target space modes. 
$\chi_i$ denotes the characters that sum up the contribution of one primary field plus its descendants.
These  characters depend on the complex modular parameter $\tau$ that parametrizes the shape of the torus,  i.e.  its complex structure.  Two tori are equivalent if their modular parameter $\tau_1$ and $\tau_2$ are connected by a modular transformation, and thus the partition function should be invariant under such modular transformations. It can be shown that the characters rotate into each other under a modular transformation, e.g. for a modular S transformation $\chi_i(-{1\over \tau}) = \sum_j S_{ij} \chi_j(\tau)$ where $S_{ij}$ is a symmetric and unitary matrix. Therefore $Z$ is invariant if $[M, S] = 0$ giving a highly non-trivial restriction on $M$ which is hard to satisfy for $M \neq 1$. To summarize the first part: When knowing $M$ we can directly state the full target space spectrum but finding non-trivial $M$ is a hard task.

Schellekens and Yankielowicz \cite{Schellekens:1989am,Schellekens:1989dq}  found a mechanism to produce non-trivial matrices $M$ in rational conformal field theories. The only object to be specified is a so called simple current $J$, whose operator product expansion with any field $\phi_i$ gives exactly one other field (plus its derivatives etc.). In terms of the fusion algebra this is written as $[J] \times [\phi_i] = [\phi_{J(i)}]$. Such simple currents actually appear in nearly all RCFTs. Having finitely many fields in a CFT implies that at some point $J^{N} = 1$, such that the application of a simple current arranges  all fields into orbits $\{ \phi, J\phi, J^2, \phi, \dots, J^{N-1} \phi \}$. 

Now, if some technical assumptions are satisfied, one can write down a matrix $M(J)$ for each simple current which gives rise to a modular invariant partition function $ Z(\tau, \ov \tau) = \vec \chi^T(\tau) M(J)  \vec \chi(\ov \tau) $. Roughly speaking $M(J)$ connects a field from the right to  fields from its orbit on the left that satisfy an additional condition involving the so-called monodromy charge. For instance space-time supersymmetry is usually implemented  by the GSO projection onto even world-sheet fermion number. 
Having more than one simple current $J_a$ any combination of their matrices $M(J_a)$ in a partition function like $Z \sim \vec \chi^T M(J_1) M(J_2)\dots  \vec \chi $ still defines a modular invariant. The order in which these matrices appear in $Z$ is important since they do not necessarily commute. In particular, states that were projected out by the first of the simple currents might enter in the twisted sector of the second simple current.

\subsection{Gepner Models}
 Gepner defined  a class of $N=2$ SCFTs describing string theory compactified on  Calabi-Yau backgrounds at  special points in their moduli space \cite{Gepner:1987qi}. In the following we review this construction and the construction of Calabi-Yau manifolds using weighted projective spaces. 

Let us split the ten-dimensional space-time into $D$ non-compact directions and $10-D$ internal directions. Using light-cone gauge effectively eliminates two dimensions so that  the external $D$-dimensional space is described by  $D-2$ free bosons. Their supersymmetric fermionic partners transform as vectors under the little group $SO(D-2)$ giving the $\widehat{\frak{so}}(2)_1$ conformal field theory. For the internal space one uses tensor products of  the minimal unitary  $N=2$ superconformal field theories (SCFT). 
The latter are exactly solvable SCFTs labeled by an integer $k>0$ admitting only a discrete set
of central charges
\eq{
c = {3k \over k+2} \, . 
} 
The primary fields  of a minimal model are labeled by three numbers $(l\; m\;s)$ with $0 \leq l \leq k$,
$-k+1\le m\le k+2$ and $-1\le s\le 2$.
A large class of  modular invariant partition functions of the minimal models are  known and follow an ADE classification. Note that in particular the D-invariant can be implemented by choosing the simple current $(k,0,0)$. To get the correct central charge of the internal space $c = 3, 6, 9$ for $10-D=2,4,6$ internal dimensions,  one has to take the  tensor product of several minimal models. In total a state in a Gepner model is
labeled as
\eq{
(l_{1} \; m_1 \; s_1) \ldots (l_r \; m_r \; s_r ) (s_0) \;\;  \in \;\; \bigotimes_{i=1}^r (k_i) \otimes \widehat{\mathfrak s \mathfrak o} (D-2)_1 \, ,
}
where we neglected the bosonic part.
Having these building blocks one needs further projections which can be taken care of by simple currents. One of them is the GSO projection which was already explained at the end of the previous section. Additionally, one needs one more simple current $J_i$ for each minimal factor ensuring that there are only pure NS and R sectors on each side. The matrices $M(J_{\rm GSO})$ and the $M(J_{i})$ all commute such that their order in the partition function  does not matter. Moreover, to be able to apply the simple current technique, one starts with the partition function of the bosonic string.  Then, at the very end  one maps it to the superstring by applying  the bosonic string map $\phi_{\rm bsm}$, which maps the four conjugacy classes of $\widehat{\frak{so}}(10)_1$ to the ones of $\widehat{\frak{so}}(2)_1$ according to $(O,V,S,C)\to (V,O,-C,-S)$. For more details, we
refer to the original literature.
The full partition function of the Gepner model is then given as
\eq{ \label{partitionfn}
              Z_{\rm Gepner}(\tau,\ov\tau)\sim \vec\chi^{\,T}(\tau)\, M(J_{\rm GSO})\,
              \prod_{i=1}^r M(J_i) \,\vec\chi(\ov\tau)\Big\vert_{\phi^{-1}_{\rm bsm}} \,.
}
From this, one can directly read off the target-space spectrum.
As already observed by Gepner \cite{Gepner:1987vz} and later well understood from different angles \cite{Greene:1988ut,Witten:1993yc}, the massless spectrum is that of  type II compactifications on certain Calabi-Yau manifolds. For this let us briefly review a construction of Calabi-Yau manifolds relevant here. One starts with a weighted projective space $W \mathbb C \mathbb P_{w_1, w_2, \dots , w_{r}}$ defined by the projective equivalence relation 
\eq{
(X_1, X_2, \dots, X_{r}) \sim ( \lambda^{w_1} X_1, \lambda^{w_2} X_2,  \dots , \lambda^{w_{r}} X_{r}) \,, \qquad \lambda \in \mathbb C \,,
}
on the complex coordinates $X_i$. The $w_i$ are called weights and due to this equivalence relation the weighted complex space has $r-1$ complex dimensions. Let us now look at hypersurfaces in the above space which we define by a single polynomial equation of degree $d$. For such a hypersurface in a weighted projective space we write $W \mathbb C \mathbb P_{w_1, w_2, \dots , w_{r}}[d]$. When the sum of all weights equals the degree $d$, the resulting $r-2$ complex dimensional hypersurface is a Calabi-Yau manifold. Similarly one can define complete intersection Calabi-Yau manifolds as intersection of two polynomials of degrees $d_1$ and $d_2$ for which we write $W \mathbb C \mathbb P_{w_1, w_2, \dots , w_{r}}[d_1 \; d_2]$. Similar as above, the resulting $r-3$ dimensional space is Calabi-Yau if $\sum w_i = \sum d_i$.

Suppose we are in 4D and we take the tensor product of five minimal models with  levels $k_i$. 
Then the target space interpretation of this background is given by a Calabi-Yau manifold defined by the Fermat-type constraint
\eq{ \label{Fermat}
X_1^{k_1+2} + X_2^{k_2+2}+ X_3^{k_3+2}+ X_4^{k_4+2}+ X_5^{k_5+2} = 0
}
in a weighted projective space with weights $w_i = {d \over k_i + 2}$. Here $d$ is the least common multiple of all $k_i + 2$. Let us turn to the easiest example with levels  $(3,3,3,3,3)$ which corresponds to the equation 
\eq{ \label{quintic}
X_1^{5} + X_2^{5}+ X_3^{5}+ X_4^{5}+ X_5^{5} = 0
}
in $\mathbb P_{1,1,1,1,1}[5]$, which is the famous quintic
(more general cases are discussed in \cite{Fuchs:1989pt,Fuchs:1989yv}).

Clearly the Fermat polynomial \eqref{Fermat} is only one single choice of a defining polynomial and any other degree 5 polynomial in the $X_i$ would also define a quintic but at another point in its complex structure moduli space. Thus,  small complex structure deformations correspond to small deformations of the defining polynomial. Such geometric deformation do correspond to massless states in the Gepner model.
For instance,  adding a term $\epsilon X_1^3 X_2^2$ to \eqref{quintic} is related to
the massless state
\eq{
X_1^3 X_2^2 \quad \sim \quad (3 \; 4 \; 1)(2 \; 3\;1)(0\;0\;0)(0\;0\;0)(0\;0\;0) (S) \,,
}
from which the general pattern should be obvious. Every Ramond ground state $(l_i\; l_i+1\; 1)$ in a minimal factor gives the power of the variable $X_i$ in the deformation.\footnote{Due to $0\leq l \leq k$,  this forbids powers of 4 and 5. Actually this is perfectly fine from the geometric viewpoint, as these can be eliminated by the freedom of  choosing coordinates.}
Therefore the combinatorics of the massless states of the Gepner model is absolutely the same as the combinatorics of the complex structure deformations.
It is important to keep in mind that all length scales in a Gepner model are of order $\alpha'$,
which often gives rise  to extra gauge symmetry enhancements.

\subsection{Gauged Supergravities}
Usually, the scalar manifold of a supergravity theory admits  a large number of isometries that can be gauged. Gauging an isometry induces couplings that can give a mass to the fields via the super Higgs effect \cite{Cremmer:1978hn,Cremmer:1978iv,Deser:1977uq,Andrianopoli:2002rm,DAuria:2007axr}. Furthermore, in $4D$ with ${\cal N} =1$ supersymmetry there exists a superpotential that  can give even more scalars a mass. When some gravitinos get a mass, as well, also the amount of supersymmetry is reduced. This breaking is in general severely constrained since e.g. the total number of degrees of freedom should be preserved.

Consider  for instance maximal ${\cal N} = 8$ supergravity in $4D$ which has only the gravity-multiplet.
Since there does not exist a superpotential,  all gaugings lead to  masses that can only come from a super Higgs effect. The field content of ${\cal N} = 8$ supergravity is precisely such that it can be decomposed into a ${\cal N} = 6$ gravity-multiplet plus two ${\cal N} = 6$ massive gravitino-multiplets.\footnote{Recall that the only massless ${\cal N}=6$ multiplet in $4D$ is the gravity-multiplet.} As such it is at least kinematically possible to break ${\cal N} = 8$ supergravity down to ${\cal N} = 6$ by gauging isometries.

But when trying to decompose the ${\cal N} = 8$ gravity-multiplet in terms of ${\cal N} = 5$ multiplets one sees that there is no way to achieve that. Therefore a breaking from ${\cal N} = 8$ to ${\cal N} = 5$ is not possible. When looking at a possible breaking of ${\cal N} = 8$ to ${\cal N} = 4$ things get more interesting. The ${\cal N} = 8$ gravity-multiplet contains 28 vectors so that  6 of them go into the ${\cal N} = 4$ gravity-multiplet and $16$ vectors go into the four massive gravitino-multiplets. The remaining 6 vectors can either form 6 massless vector-multiplets or also get a mass through additional gaugings. But since the massive vector-multiplet contains two vectors, only pairs of vectors can get a mass. When breaking ${\cal N} = 8$ to ${\cal N} = 4$ the massless spectrum is therefore ${\cal G}_{(4)} + (6 - 2k) {\cal V}_{(4)}$ where $k = 0,1,2,3$ counts the number of massive vector-multiplets.

Either by a glance into the literature \cite{Andrianopoli:2002rm}
 or by consulting a list of the supersymmetry multiplets (e.g. the appendix of \cite{Blumenhagen:2016rof}) all other breaking patterns can be easily deduced analogously to the above example. Clearly when a superpotential is induced by the gaugings/fluxes  the situation  gets a little more involved. In our case the only breaking of this kind is ${\cal N} = 2$ to ${\cal N}=1$ in $4D$, that occurs  when perturbing a type II string Calabi-Yau compactification with additional fluxes. The existence of this kind of breaking was unclear for some time but finally worked out explicitly in \cite{Louis:2009xd,Louis:2010ui,Hansen:2013dda}. When compactifying IIB on a Calabi-Yau with Hodge numbers $h_{11}$ and $h_{12}$ one gets a ${\cal N} = 2$ supergravity with $N_V = h_{12}$ vector-multiplets and $n_H = h_{11}+1$ hyper-multiplets. After breaking to ${\cal N}=1$ one can derive  the following constraints on the massless ${\cal N} =1$ spectrum
\eq{ \label{constraints}
h_{12} - h_{11} - \Delta &\leq N_V \leq h_{12} - 1 \,, \\
N_V - N_{\rm ax} &\leq h_{12} - h_{11} - \Delta \,, \\
N_V - 2 N_{\rm ax} &\geq h_{12} -2h_{11} -\Delta \,, \\
N_0 &\leq h_{12} + h_{11} \, .
}
Here $N_V$ is the number of vector-multiplets, $N_{\rm ax}$ is the number of axionic R-R multiplets while $N_0$ is the number of NS-NS scalars. $\Delta$ is either 1 or 0 depending on whether there is a gauging along the scalar that is Hodge-dual to the $B$ field. It is  important  to note  that this breaking is only possible if geometric and non-geometric gaugings are turned on simultaneously. Finding a SCFT to such a GSUGRA would therefore be  evidence, that the non-geometric fluxes/gaugings are  part of the string landscape.

\section{Asymmetric Gepner Models and their Corresponding GSUGRA}
\label{sec_corr}

Recall that we are looking for evidence for  a connection between asymmetric SCFTs and bosonic gaugings in a supergravity. The main objective  of the papers \cite{Blumenhagen:2016axv,Blumenhagen:2016rof} was to analyze this question in the framework of asymmetric Gepner models. 
While there are other  methods \cite{Israel:2013wwa}, the easiest way to construct an asymmetric Gepner model is the simple current method \cite{Schellekens:1989wx,Blumenhagen:1995tt,Blumenhagen:1995ew,Blumenhagen:1996vu,GatoRivera:2010gv,GatoRivera:2010xn,GatoRivera:2010fi,Israel:2015efa}. 
More concretely we added one or more simple currents to the partition function of the Gepner model \eqref{partitionfn} such that the full partition function becomes
\eq{
\label{gepnerasym}
              Z_{\rm ACFT} (\tau,\ov \tau)\sim \vec\chi^{\,T}(\tau)\, \underbrace{M(J_{\rm 1})\dots M(J_{\rm n})}_{\rm additional \; simple \; currents} \;
              M(J_{\rm GSO})\,
              \prod_{i=1}^r M(J_i) \,\vec\chi(\ov\tau)\Big\vert_{\phi^{-1}_{\rm bsm}} \,.
}
On the right there are still the already familiar  simple currents $J_{\rm GSO}$ and $J_i$ that guarantee that  the right-moving sector is supersymmetric. When no additional simple currents are added the supersymmetric spectrum from the right couples symmetrically to the left-moving fields. But by adding further simple currents that do not commute with $J_{\rm GSO}$ and/or $J_i$ we can enhance or break the supersymmetry on the left. Thus we still have the original supersymmetry on the right but something different on the left. In this way we can design an SCFT with  an asymmetric spectrum. Since we always have some supersymmetry from the right, we always have at least ${\cal N} = 1$ supersymmetry.  Let us now discuss a couple of simple examples.

\subsection{${\cal N}= 2$ to ${\cal N} = 1$ in $4D$}

Let us take a look at the easiest Gepner model with levels $(3,3,3,3,3)$ encountered before. As explained, it corresponds to the  quintic Calabi-Yau $\mathbb P_{1,1,1,1,1}[5]$ at the Fermat point in moduli space. Let us extend it by the simple current 
\eq{
J_{\rm ACFT} =(0\;5\;1)(0\;0\;0)^4\,(c)\,,
}
where $c$ denotes an anti-spinor representation of $\widehat{\frak{so}}(10)_1$. We notice that this simple current does not commute with the usual simple currents of the Gepner model $J_{GSO}$ and $J_i$. Being placed to the left in \eqref{gepnerasym}, it acts differently on the left- and right-movers. Hence we have an asymmetric CFT that breaks half of the supersymmetry leaving ${\cal N} = 1$ supersymmetry with the massless spectrum $(N_V, N_{\rm ax} ; N_0) = (80,0;74)$. (The notation is explained after \eqref{constraints}.) 

We want to check whether we can  interpret this asymmetric SCFT model as a GSUGRA where supersymmetry  is broken from ${\cal N} = 2$ to ${\cal N} = 1$ due to fluxes. The first step is to figure out the starting point,
which means the underlying Calabi-Yau threefold. Since we have added a simple current,
this is not expected to be the quintic. For this purpose, we have a closer look at the massless
vectors listed in table \ref{table_B1}.

\begin{table}[ht]
\centering
\renewcommand{\arraystretch}{1.2}
\begin{tabular}{|c|c|c|}
  \hline
   state & polynomial rep.  & degeneracy\\
 \hline\hline
  $(0\;1\;1)(3\;4\;1)(2\;3\;1)(0\;1\;1)^2(s)$ & $x_i^3\, x_j^2$ &
  $12$\\
$(0\;1\;1)(3\;4\;1)(1\;2\;1)^2(0\;1\;1)(s)$ & $x_i^3\,
x_j\, x_k$ & $12$\\
$(0\;1\;1)(2\;3\;1)^2(1\; 2\; 1)(0\;1\;1)(s)$ & $x_i^2\,
x^2_j\, x_k$ & $12$\\
$(0\;1\;1)(2\;3\;1)(1\; 2\; 1)^3(s)$ & $x_i^2\,
x_j\, x_k\, x_l$ & $4$\\
\hline
$(1\;2\;1)(3\;0\;0)(0\;0\;0)^3(s)+$ &  & \\
$(2\;3\;1)(3\;4\;1)(0\;1\;1)^3(s)$ & $x_i^3\, y_m$ &
  $2\times 4=8$\\
$(1\;2\;1)(2\;0\;0)(1\;0\;0)(0\;0\;0)^2(s)+$ &  & \\
$(2\;3\;1)(3\;4\;1)(0\;1\;1)^3(s)$ & $x_i^2\, x_j \, y_m$ &
  $2\times 12=24$\\
$(1\;2\;1)(1\;0\;0)^3 (0\;0\;0)(s)+$ &  & \\
$(2\;3\;1)(1\;2\;1)^3(0\;1\;1)(s)$ & $x_i \, x_j\, x_k \, y_m$ &
  $2\times 4=8$\\
\hline
     \end{tabular} 
    \caption{\label{table_B1}  Combinatorics of the $N_V=80$ massless vectors.}
\end{table}

We  realize the structure of a weighted projective space when recalling that the first number in the triplets usually tells us about the power of the variable. First of all, since the simple current only affects the first minimal factor, the other four still have weight 1 and add up to a polynomial of degree 5. For instance from states like $(1 \; 2 \; 1)(3 \; 0\;0)(0 \; 0 \; 0)^3(s)$ we are led to conclude that the first factor now acts like a coordinate of weight 2. But from counting the number of states we see that the first factor rather behaves like two coordinates of weight 2. An educated guess is then that this model has something to do  with the 
complete intersection Calabi-Yau $W \mathbb C \mathbb P_{2,2,1,1,1,1}[5 \; 3]$. Although we do not see any of the polynomials of degree 3, we inserted the 3 to satisfy the Calabi-Yau condition. This is no surprise since due to the breaking we expect many modes to be absent.

Having a candidate for the Calabi-Yau we can now take a look whether the conditions \eqref{constraints} are satisfied. We computed the Hodge numbers of $W \mathbb C \mathbb P_{2,2,1,1,1,1}[5 \; 3]$ to be $h_{21} = 83$ and $h_{11} = 2$. Plugging this into \eqref{constraints} gives only gives 6 allowed spectra in the broken phase, concretely 
\eq{(N_V,N_{\rm ax})\in\{(80,0),(80,1),(81,0),(81,1),(82,1),(82,2) \}\, .} Comparing this to the spectrum $(N_V, N_{\rm ax} ; N_0) = (80,0;74)$ of the asymmetric Gepner model we see a perfect match! 

In \cite{Blumenhagen:2016axv} we checked many more examples in a similar case by case study. The results were very encouraging for the conjecture that these ${\cal N}=1$ asymmetric SCFTs do correspond to certain ${\cal N}=2$ GSUGRAs.

\subsection{Extended supergravity}

In the previous example the existence of the superpotential yielded a  range of possible spectra in the broken theory. In situations with more supersymmetry the existence of such a superpotential is forbidden making the analysis even clearer since mass generation can only appear through the super Higgs mechanism. As additional evidence we therefore performed an analysis of asymmetric Gepner models in 8, 6 and 4 dimensions with at least 8 supercharges \cite{Blumenhagen:2016rof}. We implemented a high performance program to add up to 4 simultaneous simple currents in a stochastic search over billions of models until no new structures appeared for a long time. Our search was fine enough to find a lot more different models than the ones in a prior stochastic search \cite{Schellekens:1989wx} where no classification and no connection to GSUGRA was established. As such we tried to classify the models and test, whether they are in accordance with the ACFT/GSUGRA conjecture. The full list of models in 8,6 and 4 dimensions appeared in  \cite{Blumenhagen:2016rof} so that here we 
restrict ourselves to  state some selected classes and  the mechanisms we used to classify all models.

Let us start with $D=8$, where we do not expect to find any gauging, as  NS-NS fluxes have three legs and are therefore not supportable by the internal geometry. Nevertheless, besides the ${\cal N} =2$ maximal supergravity corresponding to type II string theory on a $T^2$, we found another model. It has ${\cal N}=1$ supersymmetry and beyond the gravity-multiplet six vector-multiplets transforming in the gauge group $SU(2)^2$. This spectrum cannot arise from a broken ${\cal N} =2$ maximal supergravity, as it has too many vectors. Therefore, one cannot interpret it in terms of a GSUGRA. Instead, we found that it is the asymmetric shift orbifold ${T^2 \over (-1)^{F_L} S W}$ where $F_L$ denotes the left-moving fermion number and $S$/$W$ denote the shift of the momentum/winding by $\frac{1}{2}$. 
Since $F_L$ contains an action onto the space-time fermions and therefore involves the R-sector, this 
is not expected to  have anything to do with a (purely bosonic) gauging. 
It turned out that all models which we could not explain as a GSUGRA admit a description in terms of a similar orbifold, mostly containing  an action by $(-1)^{F_L}$.

In $6D$ we only want to give one ${\cal N} = 1$ model to have an example with gauge symmetry enhancement. The model appeared already in \cite{Hellerman:2002ax} and has 9 tensor-multiplets, 8 vector-multiplets and 20 hypermultiplets, making   it  anomaly free. We also found small variations of this model with up to four additional vector/hyper pairs. It turns out, that the additional vectormultiplets are very different in nature to the 8 vectormultiplets in the basic model. The original vectors appear in the R-R sector and cannot couple to the hypermultiplets. In contrast, the additional vectors arise  from the NS-NS sector and couple to the hypermultiplets. As such they develop a mass when a scalar from the hypermultiplets gets a vacuum expectation value. This looks like a gauge symmetry enhancement where the vacuum expectation value of a hypermultiplet controls the deviation from the enhancement point where additional vectors become massless. Away from the enhancement point the additional vector/hyper pairs get a mass such that we land at the basic model. Recall that being at the self dual radius such symmetry enhancements are not unexpected  for Gepner models.

Let us turn to the $4D$ case. Again we refrain from  presenting the complete result, but concentrate  on a selection that gives a good overview over our methods. Here are some of the classes we found:
\begin{itemize}
\item ${\cal N}=8$ maximal supergravity which is type II on a $T^6$.
\item ${\cal N}=6$ supergravity with its gravity-multiplet. As described before, this can be interpreted as broken ${\cal N} = 8$ SUGRA. Alternatively, there exists an orbifold description of this model given by ${T^6 \over \mathbb Z_2^L S}$ where the $\mathbb Z_2^L$ is a reflection on four of the left-moving coordinates and $S$ denotes a shift along the other two coordinates.
\item ${\cal N} = 5$ supergravity with its gravity-multiplet. As mentioned, this cannot arise as a broken ${\cal N} = 8$ supergravity. Instead, it can be realized as orbifold ${T^6 \over \mathbb Z_2^L S, \tilde{\mathbb Z}_2^L \tilde S }$ with orthogonally shifted asymmetric reflections. 
\item ${\cal N} = 4$ supergravity with $N_V = 0,2,4,6,8,10,14,18$ vectormultiplets besides  the gravity-multiplet. Recall from above that ${\cal N}=8$ can break to ${\cal N} = 4$ supergravity with $N_V = 0,2,4,6$ vectormultiplets, so only part of the spectrum can be explained by a gauging. They instead can be explained by a similar orbifold as the one appearing in $8D$ with its $SU(2)^2$ gauge group. The $4D$ model was found in \cite{Dixon:1987yp} and has gauge group $SU(2)^6$ and therefore explains the upper bound of the models we found. The other models $N_V = 6,8,10,14$ are then naturally explained as the Coloumb branch of this orbifold. Notice that the model $N_V =6$ can be seen both as Coloumb branch or gauged ${\cal N} = 8$ supergravity and that we have small holes in the spectrum $N_V = 12,16$ which might disappear in an even more detailed scan. 
\item ${\cal N} = 3$ supergravity with $N_V = 3,7,11,13,19$ vectormultiplets besides the gravity-multiplet. These models were rather rare, so we suspect the full list to have all odd numbers of vectormultiplets. These models can perfectly be explained by GSUGRA. Indeed, breaking the maximal ${\cal N}=8$ supergravity down to ${\cal N} = 3$ gives $3$ or $1$ vectormultiplets. When breaking the ${\cal N} = 4$ supergravity of a type II on $K3 \times T^2$ down to ${\cal N} = 3$ the spectrum is constrained to $N_V = 19-2k$ where $k \in \mathbb N_0$. The data we found are therefore perfectly consistent with the super Higgs effect. 
\end{itemize}

\noindent
Let us stop here and refer to the paper \cite{Blumenhagen:2016rof} for the full list that also includes 
 ${\cal N} = 2$  vacua. We were able to provide  similar interpretations to all examples we found.
Either the models were completely consistent with the super Higgs effect
or could be explained by some other mechanism.
Thus,  we did not encounter any contradiction  to the ACFT/GSUGRA conjecture in the billions of models we checked.

\vspace{1cm}

\noindent
\emph{Acknowledgments:} This article summarizes two talks given at the  ``Workshop on Geometry and Physics'', which took place on
November 20-25, 2016 at Ringberg Castle, Germany. The workshop was dedicated 
to the memory of Ioannis Bakas, a fine colleague who will be remembered 
for his commitment and enthusiasm for  physics and the many interesting discussions we had. 
It is a great loss.



\end{document}